\documentclass[useAMS,usenatbib]{mn2e}
\usepackage{psfig,morefloats,url,amsmath}
\usepackage{graphicx}
\usepackage{subfig}

\footnotesize
\newdimen\digitwidth    %define ! a one digit width for tables
\setbox0=\hbox{\rm0}
\digitwidth=\wd0
\catcode`!=\active
\def!{\kern\digitwidth}

\normalsize

\title[Galactic centre pulsar detection]{Detecting pulsars in the Galactic centre}

\author[K. Rajwade et al.]
  {K. M.~Rajwade$^{1,2}$\thanks{kmrajwade@mix.wvu.edu},
  D. R.~Lorimer$^{1,2,3}$ and L. D.~Anderson$^{1,2,3}$\\
  1.~Department of Physics and Astronomy, West Virginia University, 
  Morgantown, WV 26506, USA\\
  2.~Center for Gravitational Waves and Cosmology, West Virginia University,
  Chestnut Ridge Research Building, Morgantown, WV 26505, USA\\
  3.~Green Bank Observatory, Green Bank, WV 24944, USA}
  
\date{\today}

\pagerange{\pageref{firstpage}--\pageref{lastpage}} \pubyear{2002}

\def\LaTeX{L\kern-.36em\raise.3ex\hbox{a}\kern-.15em
    T\kern-.1667em\lower.7ex\hbox{E}\kern-.125emX}

\begin{document}

\label{firstpage}

\maketitle

\begin{abstract}
Although high-sensitivity surveys have revealed a number of highly
dispersed pulsars in the inner Galaxy, none have so far been found in
the Galactic centre (GC) region, which we define to be within a
projected distance of 1~pc from Sgr~A*. This null result is surprising given that several independent lines of evidence 
predict a sizable population of neutron stars in the region. Here, we present a detailed analysis of both
the canonical and millisecond pulsar
populations in the GC and consider free-free absorption and multi-path
scattering to be the two main sources of flux density mitigation. We
demonstrate that the sensitivity limits of previous surveys are not
sufficient to detect GC pulsar population, and investigate the optimum
observing frequency for future surveys. Depending on the degree of
scattering and free-free absorption in the GC, current surveys constrain
the size of the potentially observable population (i.e.~those beaming towards us) to be
up to 52 canonical pulsars and 10000 millisecond pulsars. 
We find that the optimum frequency for 
future surveys is in the range of 9--13~GHz. We also predict that future deeper
surveys with the Square Kilometer array will probe a significant
portion of the existing radio pulsar population in the GC. 
\end{abstract}


\section{Introduction}
Understanding the stellar populations in the Galactic centre (GC)
region, and how they relate to the central super-massive black hole
(Sgr A*), is a major goal of modern astrophysics. The central few
parsecs of the Galaxy are known to consist of large molecular
complexes and have high stellar densities compared to the rest of the
Galactic disk~\citep[see, e.g.,][]{Sc07}.  Under these circumstances,
many authors have already made the reasonable assumption that a large population
of neutron stars exist in the GC~\citep{Mo96,Ge10}.

Motivated by the promise of finding pulsars orbiting Sgr A*, there
have been multiple surveys of the GC
region~\citep{Jo06,Ma10,De10,Ba11}.  These surveys are typically
conducted at frequencies higher than $\sim$1~GHz to reduce the impact
of interstellar scattering, which is known to cause potentially significant pulse broadening along
lines of sight to pulsars in the inner Galaxy~\citep{Co97}. To date, no pulsars have been found in the GC region which we define in this paper to be within 1~pc of Sgr~A* (i.e.~an angular offset
of 25$^{"}$ for $R_{0} = 8.3$~kpc). The discovery of a magnetar~\citep{Ea13,Mo13} has
brought the problem of pulsars in the GC to fore again. The discovery 
led~\cite{Ch14} and \cite{De14a} to conclude that there are very few pulsars in 
the GC. Moreover,~\cite{De14a} also go on to suggest that the magnetar 
formation efficiency is higher in the GC and the lack of detection could be attributed to shorter life spans of magnetars. These results seem
puzzling, given the high stellar density of the GC. Also, angular
broadening measurements of SGR~1745$-$2900 have revealed that the
scattering along this line of sight is less than
expected. \cite{Bo14} claim that the scattering screen along the line
of sight lies $\sim 6$~kpc away from the GC in the Scutum 
arm. These findings suggest that previous surveys should have discovered more pulsars in the GC, which makes their dearth baffling.

The presence of hot, ionized gas in the central parsec of our Galaxy~\citep{Pe89,Gi12} 
raises the question of whether absorption can affect
detection of radio pulsars. Recent studies have shown
free-free thermal absorption to be the primary source of gigahertz
peaked spectra, where the flux density spectrum shows a turnover at frequencies of $\sim$1~GHz in some pulsars found in dense ionized
environments~\citep{Le15,Ra15}. The GC environment suggests that absorption 
by ionized gas could decrease the observed flux density from neutron stars beaming 
towards us.

Such a dense and highly turbulent environment can also be responsible for large 
scattering, thereby, reducing incoming pulsar radio flux density in our line of sight. 
The effects of the interstellar medium (ISM) in the GC on pulsar flux densities have
been studied previously.~\cite{Co97} modeled multi-path scattering in
the GC in terms of a thin screen near the centre. As a result, the
radio pulses observed can be substantially broadened at lower
frequencies. \cite{Wh12} studied various flux density mitigation effects due to 
the ISM that can alter the incoming pulsar flux and result in a
non-detection. Recently, \cite{Ma15} proposed that the neutron star
population of the GC is dominated by millisecond pulsars (MSPs). They
also claimed that more sensitive, high frequency surveys in the future
would be able to detect MSPs in the GC. Though a MSP population has
been predicted in the past, the results of \cite{Ma15} are based on
the pseudo-luminosity distribution of known pulsar population sample, which has an 
inherent pseudo-luminosity bias since we only detect the brightest pulsars.

In this paper, we try to answer questions regarding the GC pulsar population 
by modeling the GC environment and accounting for observational selection 
biases. We simulate a pulsar population in the GC environment and study the 
effect of the GC environment on pulsar flux densities. We find the optimum frequency 
for future surveys based on the results of the simulation. Section 2 describes 
the simulations with different models considered. In Section 3, we present the 
results of the analysis and their implications.

\section{Simulations}

To place constraints on the number of pulsars in the GC, we simulated synthetic populations 
of pulsars using the PsrPopPy package~\citep{Ba14}, a python module based on the {\tt psrpop}
code developed earlier for population synthesis of pulsars~\citep{Lo06}. The inferred parameters 
from the known pulsar population in the Galaxy are biased due to various selection effects~\citep[see, e.g.,][]{fk06}. These effects are accounted for by PsrPopPy~\citep[see][for details]{Ba14}. PsrPopPy generates synthetic pulsar populations based on a set of pulsar parameters. These are then
searched for in a simulated pulsar survey based on past survey parameters to determine the 
subset of pulsars, that are theoretically detectable.  

We considered populations of canonical pulsars (CPs) and millisecond pulsars (MSPs) in our 
analysis with PsrPopPy (Bates et al.~2014). 
For both cases, we simulated the populations using the pseudo-luminosity 
scaling with period and period derivative. Following previous authors, we 
parametrize the 
pseudo-luminosity $L$ in terms of period $P$ and period derivative $\dot{P}$ as a power law:
\begin{equation}
  L = \gamma P^{\alpha} \dot{P}^{\beta},
\end{equation}
where $\alpha$, $\beta$ and $\gamma$ are model parameters. For simplicity,
following~\cite{Ba14}, we take $\alpha=-1.4$ and
$\beta=0.5$ which physically links $L$ to be proportional to the square
root of the pulsar's spin-down pseudo-luminosity. The uncertainties 
on $\alpha$ and $\beta$ are reported in~\cite{Ba14}. To ensure that errors 
on $\alpha$ and $\beta$ do not affect our results, we reran our simulations 
by changing one parameter by 1$\sigma$ and kept the other same. We observed 
that changing the parameters within the errors had little to no effect on the 
results as discussed later. To ensure that the properties of the 
simulated sample are comparable to the observed sample, we modified the 
constant of proportionality in this expression, $\gamma$ so that
the pseudo-luminosity of the simulated 
sample that is detected in a simulated Parkes survey matches the observed 
detected sample in the same survey, assuming that the properties of the pulsars in the GC are similar to the properties of detected pulsars.
To achieve this, we simulated a population of CPs and MSPs 
for different $\gamma$s and ran a Kolmogorov-Smirnov (K-S) test on the 
pseudo-luminosity distributions of the simulated and the observed sample for both 
sub-populations. Since the K-S probability beyond $\sim$~0.1 implies that the model
and observed distributions are statistically indistinguishable~\cite[see, e.g.,][]{Pr02}, 
we obtain a range of $\gamma$ values for which the luminosities are consistent as shown 
in Fig.~\ref{fig:KSprob}. The best 
$\gamma$ was chosen for the case where we obtained the maximum K-S probability 
for the two detected populations. The best simulated populations were used for further analysis. The parameters used for simulation of both populations are given in Table~\ref{tab:simparams}.
We note in passing here that the optimal values of $\gamma$ found here imply
population-averaged luminosity values of 2.1 mJy~kpc$^2$ and 0.1 mJy~kpc$^2$ for CPs and MSPs
respectively. Although our analysis does not make any distinction between solitary
and binary MSPs which appear to have different luminosities~\citep{Ba97,Bu13}, it does clearly show that MSPs are
intrinsically fainter radio sources than CPs.

\begin{table*}
\begin{tabular}{lc lc lc }
\hline
Parameter & CP & MSP  \\
\hline \hline
Radial distribution Model & \cite{Lo06} & \cite{Lo06}\\
\\
Initial Galactic $z$-scale height & 50~pc & 50~pc \\
1-D velocity dispersion  &  265~km~s$^{-1}$    &   80~km~s$^{-1}$   \\
Maximum initial age & 1~Gyr & 5~Gyr \\
\\
Luminosity parameter $\alpha$ & -1.4 (1) & -1.4 (1)\\
Luminosity parameter $\beta$ & 0.50 (4) & 0.50 (4) \\
Luminosity parameter $\gamma$ & 0.35 & 0.009 \\
\\
Spectral index Distribution & Gaussian & Gaussian \\
$\langle \alpha \rangle$ & -1.4 & -1.4 \\
$\sigma_{\alpha}$ & 0.9 & 0.9 \\
\\
Initial Spin period distribution & Gaussian & Log-Normal~\citep{Lo15}\\
$\langle P \rangle$~(ms)  & 300  & --- \\
$\sigma_{P}$ (ms) &150 & --- \\
$\langle \log_{10}{\rm P(ms)} \rangle$  &--- & 15 \\
$std(\log_{10}{\rm P(ms)})$ &--- & 56 \\
\\
Pulsar spin-down model & \cite{fk06} & \cite{fk06}\\
\\
Beam alignment model & orthogonal & orthogonal \\
\\
Braking Index & 3.0 & 3.0 \\
\\
Initial B-field distribution & Log-normal & Log-normal \\
$\langle \log_{10}{\rm B(G)} \rangle$ &12 & 8 \\
$std(\log_{10}{\rm B(G)})$ & 0.55 & 0.55\\
\\
Observed sample size & 1065 & 39\\
\hline
\end{tabular}
\caption{Table showing the different model parameters used in PsrPopPy for simulation of the two pulsar populations. The values used in the simulation are adopted from previous Parkes surveys~\citep{Ma01,Ed01}. Values in the parenthesis indicate 1-$\sigma$ uncertainties on the least significant digit.}
\label{tab:simparams}
\end{table*}

\begin{table*}
\begin{tabular}{lc lc lc lc lc lc lc}
\hline
Survey & Frequency &T$_{\rm sys}$& t$_{\rm int}$ & G & S/N$_{\rm min}$ & $\Delta\nu$  \\
\hline
 & (GHz) & (K) &(s) & (K~Jy$^{-1}$) &  & (MHz) & \\ 
\hline
Bates et al. 2011 &6.5 &40 &1055 &0.6 &10 &576 \\
Macquart et al. 2010 & 15&35 &21600&1.5&10 &800  \\
Johnston et al. 2006 & 8.4&40 &4200 & 0.6& 10&864  \\
SKA-MID &5 &30 &50400 &17.7 &10 &770 \\
ngVLA &10 &34 &25200 &22.4 &10& 8000 \\
\hline
\end{tabular}
\caption{Basic parameters for previous and future pulsar surveys towards the GC.}
\label{tab:survconf}
\end{table*}

\begin{table*}
\begin{tabular}{l|rr|rr|rr|rr|rr}
Model         & \multicolumn{10}{c}{Survey} \\
         &    \multicolumn{2}{c}{A} & \multicolumn{2}{c}{B} &
              \multicolumn{2}{c}{C} & \multicolumn{2}{c}{D} &
              \multicolumn{2}{c}{E} \\
\hline
 & CP & MSP & CP & MSP & CP & MSP & CP & MSP & CP & MSP \\
\hline
 BL  & 353  &  4840 & 444  & 5200 & 130  &  820 & 11  & 38  & 10 & 27 \\
 WS  & 354 &  4950 & 444  & 5775  & 130  &  830 & 11  & 44 & 10 & 27 \\
 SS  & 375  &  --  & 445  &  --  & 130  &  1020 & 13  &  -- & 10 & 350 \\
 FF  & 353  & 4900 & 444  &  5700 & 130  &  820 & 11  & 41 & 10 & 27 \\
 FF+WS  & 360  &  5585 & 444  & 5800 & 130  & 830 & 12  & 44 & 10 & 27 \\
 FF+SS  & 378  &  --  & 445  &  --  & 130  &  1020 & 13 &  -- & 10& 350 \\
\hline
\end{tabular}
\caption{Table showing 95$\%$ confidence level upper limits on the population for a null result in previous and future surveys. These are conservative limits since we use the lowest acceptable $\gamma$ values. The surveys considered here
are: (A) Bates et al.~(2011); (B) Johnston et al.~(2006);
(C) Macquart et al.~(2010); (D) SKA-MID survey; (E) ngVLA survey.
The models listed are: (1) the baseline (BL) model with no scattering or free-free absorption; (2)
weak scattering (WS); (3) strong scattering (SS); (4) free-free absorption (FF);
(5) free-free and weak scattering (FF+WS); (6) free-free and strong scattering (FF+SS). 
\label{tab:survres}}
\end{table*}

We scaled the derived luminosities of the simulated population at 1.4~GHz to different frequencies given in Table~\ref{tab:GCsurv} for both populations assuming a normal distribution of spectral indices~\citep{Ba13}. Then, the corresponding observed flux density
\begin{equation}
S = \frac{L_{\nu}}{D_{\rm GC}^{2}},
\label{eq:flux}
\end{equation}
where $\rm L_{\nu}$ is the pseudo-luminosity at a frequency $\nu$~\citep[see][for details]{Ch14} and 
$D_{\rm GC}$ is the distance to the GC which is assumed to be 8.3~kpc~\citep{Bo14}.

\begin{figure*}
\centering

%\centerline{\psfig{file=Psrpop_surv_cp_t1.eps,width=15 cm}}
%\centerline{\psfig{file=Psrpop_surv_msp_t1.eps,width=15 cm}}
\includegraphics[scale=0.3]{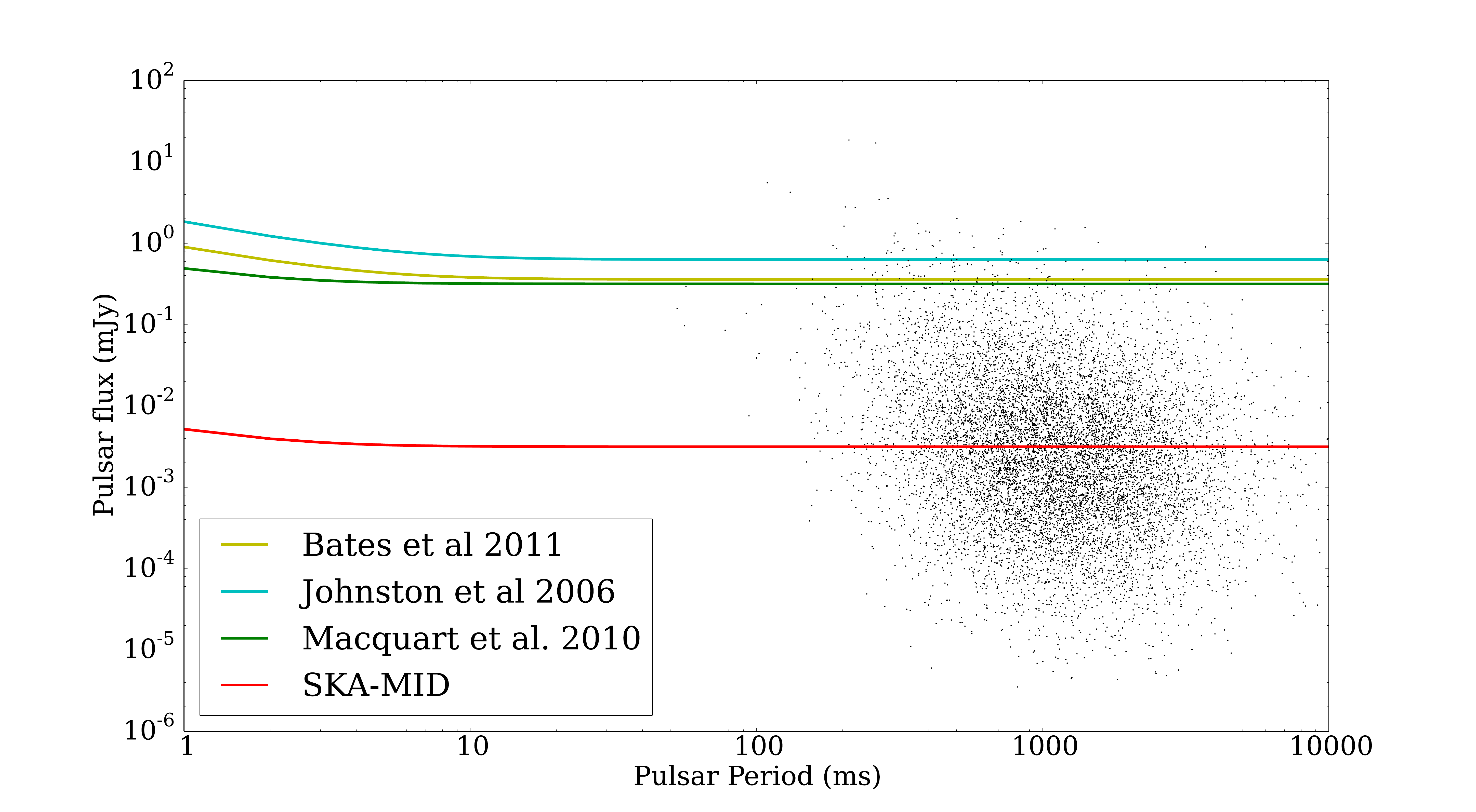}
\\
\includegraphics[scale=0.3]{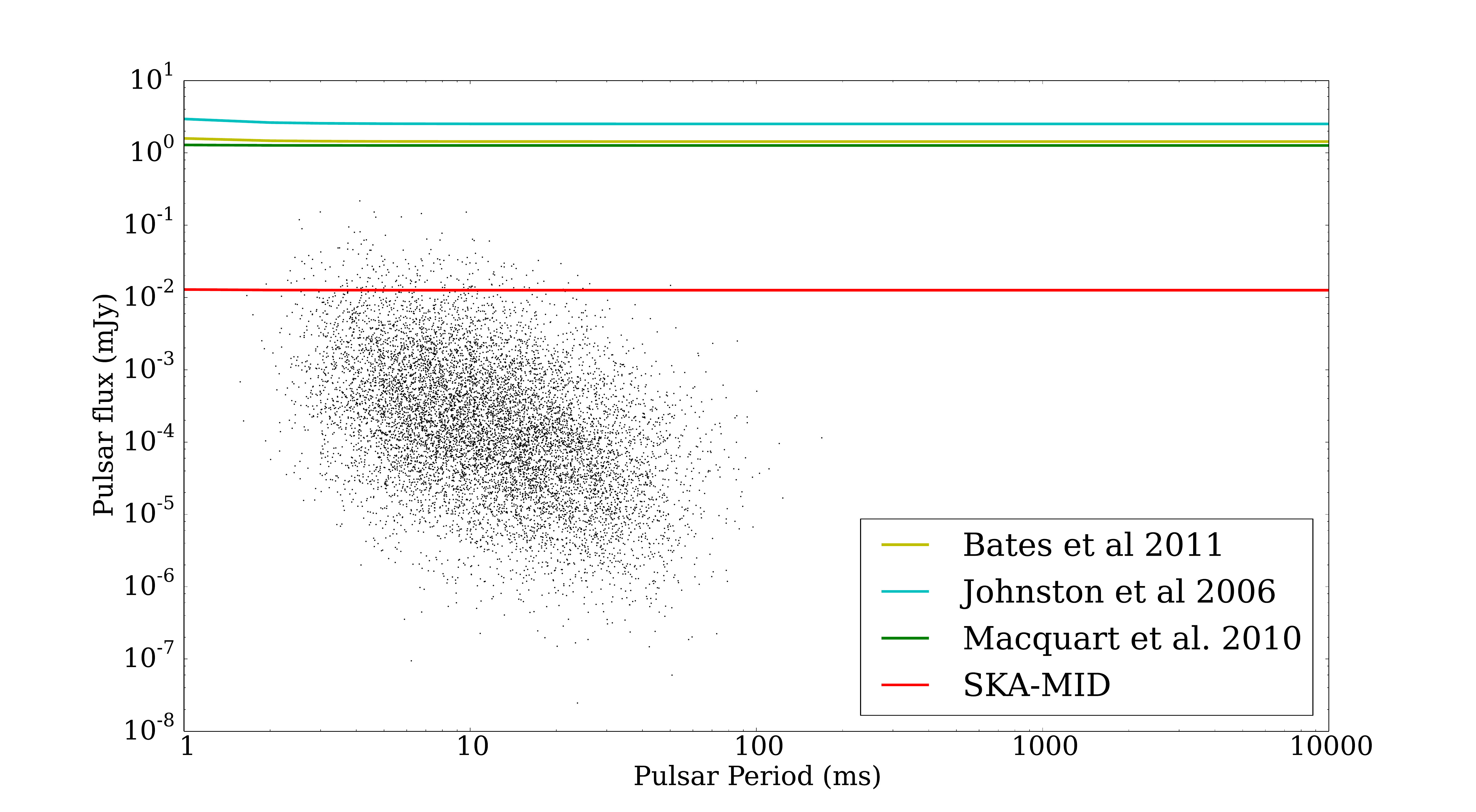}

%\end{tabular}
\caption{1.4~GHz mean flux density versus period for a synthetic
   population of 10000 pulsars at the GC for the baseline model (BL). The top panel shows CPs while the 
bottom panel shows MSPs.  Different lines indicate the survey sensitivities of
  past surveys apart from the SKA-MID survey. The parameters for SKA-MID survey are the expected parameters of the telescope. The sensitivities of each survey have been
  scaled to 1.4~GHz assuming a spectral index of --1.4~\citep{Ba13}. The flux density limit curves for each survey correspond to a DM of 1780~cm$^{-3}$~pc. The flux sensitivity limit for ngVLA is not shown since it almost overlaps with the SKA-MID flux limit.}
\label{fig:surv}
\end{figure*}

We obtained flux densities for different frequencies from luminosities obtained in the simulations using
Eq.~\ref{eq:flux}. Then, using the models discussed in the subsequent 
sections, we multiplied the flux densities by the appropriate factors to account for the reduction due to three scenarios: (i) Scattering, where the flux density is reduced due to multi-path
scattering between the source and the observer; (ii) Free Free 
absorption, where the radio flux density from the pulsar is absorbed by the 
intervening medium; (iii) Both scattering and free-free absorption 
playing a role in flux density mitigation. Under these circumstances, we calculated 
the upper limit on CP and MSP populations for previous and future surveys. For 
a given model, survey and a sample size, we ran multiple realizations of our 
simulations and we repeated the same experiment for a range of sample sizes of 
the GC pulsar population. Then, we counted the number of times our simulations 
produce no detection for a given sample size. This exercise enabled us to 
generate a probability density function (PDF) of these null results as a 
function of sample size. To compute the 95$\%$ confidence level upper limit on 
the CP and MSP populations, we calculated the sample size such that the area 
under the PDF was 0.95 times the total area under the curve. For this 
analysis, where we report conservative limits on the GC pulsar populations, we used the lowest $\gamma$ value above a K-S probability of 0.1. Since the change in $\alpha$ and $\beta$ within 1$\sigma$ errorbars affected the number of 
detected pulsars in a given survey by a factor of $\sim$1, we conclude that 
the change in those parameters does not affect our upper limits. The results of this analysis are shown in Table~\ref{tab:survres}. Figure.~\ref{fig:surv} shows the baseline simulation of CPs and MSPs with past survey sensitivities overlaid along with future SKA-MID and ngVLA~\citep{Ca15} survey with assumed
parameters of the telescope~\footnote{\url{https://www.skatelescope.org/wp-content/uploads/2012/07/SKA-TEL-SKO-DD-001-1_BaselineDesign1.pdf}}. The results of
the past surveys along with the ngVLA and SKA-MID survey are shown in Table~\ref{tab:survres}. From this it is evident that, even without considering any effects of the GC environment on the pulsar
flux densities, the past surveys have been insensitive to
the total pulsar population in the GC.

\subsection{Model}

In an attempt to make sense of the lack of pulsars in the GC found so
far, we developed a model described below that takes account of
multi-path scattering and free-free absorption effects on the pulsar
signal. If $S_{\rm 0}$ is the intrinsic flux density of a pulsar at a frequency $\nu$,
 then the measured flux density at the telescope
\begin{equation}
S_{\nu} = S_{{\rm 0},\nu}~\mathcal{S(\nu)}~\mathcal{F(\nu)}, 
\end{equation}
where $\mathcal{S(\nu)}$ and $\mathcal{F(\nu)}$ are the flux density mitigation factors 
due to scattering and free-free absorption respectively.
These factors are discussed in turn in the sections below.

\subsubsection{Free-Free absorption}

Free-free absorption is known to bias flux density spectra of some
pulsars~\citep{Le15,Ra15}. This is manifested by a turnover in pulsar
spectra at frequencies of $\sim$1~GHz~\citep{Ki07,Ki11} which is
different from the turnover seen at lower
frequencies due to synchrotron self absorption~\citep{Si73}. This phenomenon is normally observed in
pulsars that lie in dense environments like pulsar wind nebulae
or supernova remnants.  Since the GC consists of dense, ionized gas
and cold molecular gas with thin ionization fronts, we assume
free-free absorption plays a part in reducing the flux density of an
expected pulsar population at the GC.  If $\tau$ is the optical depth
along a given line of sight then, as we showed in \cite{Ra15}, the observed
flux density
\begin{equation}
S_{\rm obs,\nu} = S_{\rm ref,\nu_{\rm ref}}~\left(\frac{\nu}{\nu_{\rm ref}}\right)^{\alpha}~\mathcal{F(\nu)},
\label{eq:gps}
\end{equation}
where
\begin{equation}
\mathcal{F(\nu)} = \exp\left[-\tau_{\nu} \left(\frac{\nu}{\nu_{\rm ref}}\right)^{-2.1}\right],
\end{equation}
and $S_{\rm ref,\nu_{\rm ref}}$ is the pulsar's observed
flux density at a reference frequency $\nu_{\rm ref}$ at which $\tau_{\nu} \ll$1. 
For a correction factor of order unity\footnote{This assumption is
reasonable so long as $T_e>20$~K, which is the case in this work.}, 
the optical depth
\begin{equation}
\tau_{\nu} = 0.082 \left(\frac{\nu}{\rm GHz}\right)^{-2.1} 
\left(\frac{\rm EM}{{\rm cm}^{-6}~{\rm pc}}\right) 
\left(\frac{T_{e}}{\rm K}\right)^{-1.35}.
\label{eq:tau}
\end{equation}
For this analysis, following \cite{Pe89}, we adopt
an emission measure EM = 5$\times$10$^{5}$~cm$^{-3}$~pc and 
electron temperature $T_{e}$= 5000~K for the GC.~\cite{Ra15} shows that this effect is smaller at frequencies greater than $\sim$1~GHz, which will be discussed later.

\subsubsection{Scattering}

\begin{figure}
\includegraphics[scale=0.25]{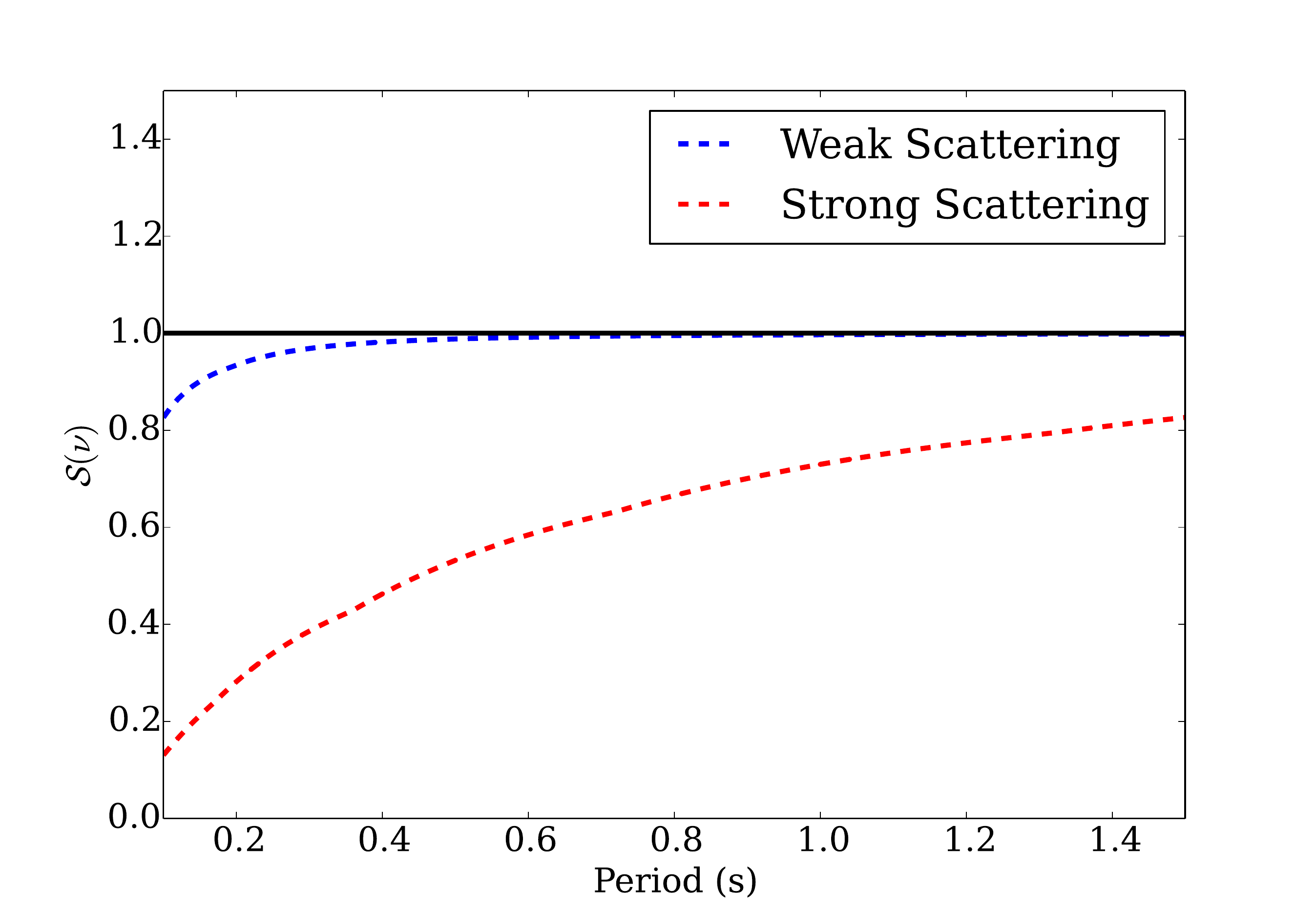}
%\end{tabular}
\caption{Scattering efficiency ($\mathcal{S}(\nu)$) as a function of period 
for CPs. The horizontal black line corresponds to $\mathcal{S}(\nu)$ = 1. }
\label{fig:eff}
\end{figure}

Given a flux density spectrum that is modified by free-free absorption
in the GC region, we also need to consider the impact of multi-path
scattering. Observations of scatter-broadened pulse profiles, which
are typically in the form of a one-sided exponential, have long been
known to be powerful probes of the physical composition and structure
of the ISM \citep[for a review, see e.g.,][]{Kr15}. Since the GC is a
region with high stellar density and large amounts of molecular and
ionized gas, a significant amount of scattering is expected for
pulsars in this region. From~\cite{Co97}, for observations at some
frequency $\nu$ and scattering due to a thin screen, the corresponding
scattering timescale
\begin{equation}
\begin{aligned}
t_{\rm sca}(\Delta_{\rm GC}) &= 6.3 {\rm s} \, \left(\frac{D_{\rm GC}}{8.5~{\rm kpc}}\right) 
\left(\frac{\theta_{\rm GC, 1~GHz}}{1.3\,''}\right)^{2} \\
&
\left(\frac{\nu}{\rm GHz}\right)^{-4}~\left(\frac{D_{\rm GC}}{\Delta_{\rm GC}}\right)~\left(1 - \frac{\Delta_{\rm GC}}{D_{\rm GC}}\right).
\end{aligned}
\end{equation}
In this expression, $D_{\rm GC}$ is the distance to the GC,
$\Delta_{\rm GC}$ is the distance of the scattering screen from the GC
and $\theta_{\rm GC}$ is the angular broadening of Sgr~A* scaled to a
frequency of 1~GHz. We compute $\mathcal{S(\nu)}$ following the 
treatment in~\cite{Co97} 
and~\cite{Co97b}. We assume pulses to be characterized by a Gaussian and convolve this with a
one-sided exponential scattering function to 
broaden the pulse. In Fourier space, the amplitude of the harmonics 
will be the product of the Fourier transform of the Gaussian pulse 
and the scattering function. Since 
the scattering reduces the peak amplitude of the pulse, that manifests itself as a reduction in the efficiency of the survey. We define this efficiency
\begin{equation}
\mathcal{S(\nu)} = \frac{\eta_{\rm p,sc}}{\eta_{\rm p,std}},
\end{equation}
where $\eta_{p,sc}$ is the pulsed fraction for the scattered pulse and $\eta_{p,std}$ is the pulsed fraction of the standard Gaussian pulse (See Appendix A for details). The position of the scattering screen towards the GC is still
uncertain. For this analysis, we assume strong scattering scenario
with the screen at $\sim$130 pc~\citep{Co97} and weak scattering with
screen at $\sim$6 kpc from the GC~\citep{Bo14}.
We 
did these calculations for CPs and MSPs for weak and strong scattering. 
Figure~\ref{fig:eff} shows the efficiency as a function of period for CPs. 
For this analysis, we used a constant duty cycle of 0.4 for MSPs and 0.05 for 
CPs.

\begin{figure*}
\centering
%{\mbox{\psfig{file=2D_Hist_v_t1_ar.eps,width=18cm,height=10cm}}}
%{\mbox{\psfig{file=B1133+16_simspec.eps,width=8cm}}}
\includegraphics[scale=0.4]{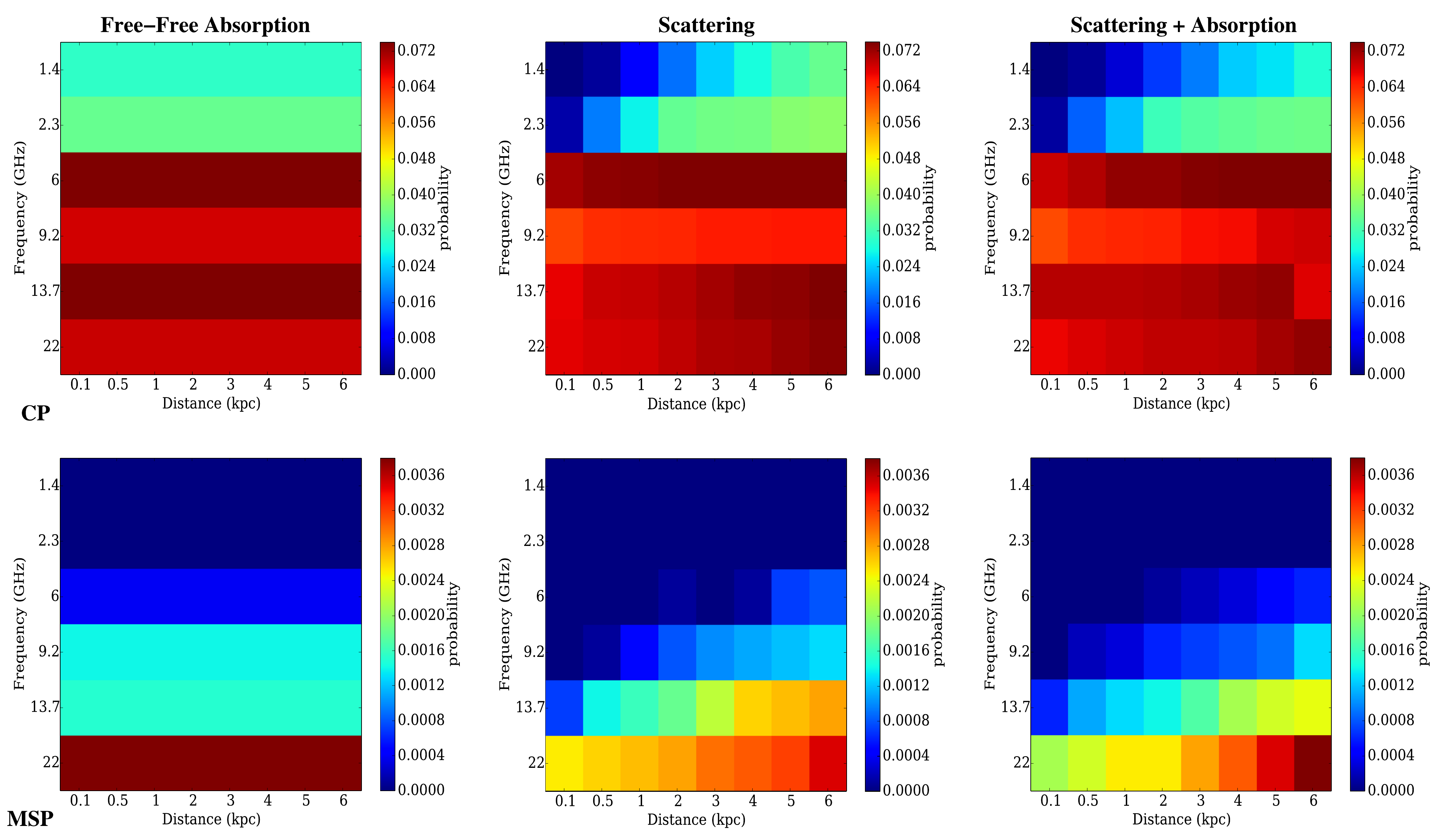}
\caption{Probability of finding a pulsar in the GC as function of frequency and distance 
of the scattering screen from the GC in future GBT surveys assuming
that the backend would be able to incorporate the whole bandwidth of each
receiver. The columns from left to right are: free-free absorption, scattering, both
scattering and absorption. The upper row is for CPs while the bottom one is for MSPs. }
\label{fig:2dhist}
\end{figure*}

\subsection{Probability of detection}

Finally, we computed a probability of detecting a single pulsar (CP and MSP) at the GC as a 
function of frequency and screen distance for each of the three scenarios (scattering, free-free
absorption and both effects) by considering surveys of the GC with the Green Bank Telescope (GBT).
We selected the GBT because it is one of the largest fully steerable 
single dish telescope where one can observe the GC for a significant duration. We adopted
the known parameters of Green Bank 
Telescope (GBT) receivers from the GBT observing guide\footnote{\url{https://science.nrao.edu/facilities/gbt/proposing/GBTpg.pdf}} to compute the flux density limit at 
different frequencies for future GBT surveys (see Table~\ref{tab:GCsurv}). The 
sky contribution from the GC to the system temperature is significant and 
since the GC transits at an elevation of $\sim$21$^\circ$, it was necessary to 
account for the changes in the system temperature, $T_{\rm sys}$ at lower elevations. To do this we assumed, the system temperature of each receiver,
\begin{equation}
T_{\rm sys}  = T_{\rm GC} + T_{\rm atm} + T_{\rm rec},
\label{eq:T}
\end{equation}
where, $T_{\rm GC}$ is the contribution of the GC, $T_{\rm atm}$ is the 
contribution due to the atmosphere and $T_{\rm rec}$  is the constant receiver 
temperature. $T_{\rm GC}$ is computed by taking the weighted average of $T_{\rm GC}(\nu)$ over the 
band of the receiver. To compute $T_{\rm GC}(\nu)$, we used the recent 
continuum maps of the GC at 1.4, 6 and 9.2~GHz from~\cite{la08}. Using the 
calibrated maps, we used the flux density at the pixel coresponding to the GC to fit a power-law which led to a relationship
\begin{equation}
T_{\rm GC}(\nu) = 568~\left(\frac{\nu}{\rm GHz}\right)^{-1.13}~\rm K.
\end{equation}
For $T_{\rm atm}$, we computed  empirical relations
between $T_{\rm atm}$ and elevation for each receiver which made use of data from 
the GBT sensitivity calculator~\footnote{\url{https://dss.gb.nrao.edu/calculator-ui/war/Calculator_ui.html}}. Then, we computed the weighted average of 
$T_{\rm atm}$ over all hour angles of the source by taking into account the dependence
of elevation with hour angle. The final $T_{\rm sys}$ is calculated by 
plugging in values for $T_{\rm GC}$, $T_{\rm atm}$ and $T_{\rm rec}$ in Eq.~\ref{eq:T}. The final values of flux density sensitivities are given in Table~\ref{tab:GCsurv}
For multi-beam receivers, we assumed 
only a single beam. In these calculations, we are not assuming 
any coherent summing of multiple epochs. Using the flux densities computed in the simulation, we 
obtained flux density histograms of the synthesized population at different GBT 
frequencies and counted up the number of the pulsars above the flux density threshold 
of each survey. The required detection probability is simply the ratio of
pulsars above each survey threshold to the total number of pulsars simulated.

\begin{table*}
\begin{tabular}{lc lc lc lc lc lc}
\hline
Receiver & Central Frequency & Bandwidth & 10-$\sigma$ Sensitivity Limit & VEGAS Limit &\multicolumn{4}{c}{Detection probabilities expressed as percentages}\\
\hline
      & (GHz) & (MHz) & $\mu$Jy & $\mu$Jy &\multicolumn{2}{c}{Future backends} &\multicolumn{2}{c}{VEGAS}  \\
 & & & & & CP & MSP & CP & MSP \\
\hline
L-Band& 1.4& 650 &119 & 119 &$\leq 3.5$ &0.0 &$\leq 3.5$ & 0.0  \\
S-Band& 2.3& 970 &62.3 & 62.3  &$\leq 3.9$ &0.0 & $\leq 3.9$ &0.0 \\
C-Band & 6 & 3800 &12.2 & 20.3 & 8 &0.08 & 5.3 &$\leq 0.04$ \\
X-Band & 9.2 &2400 &11.3&16.3 & 7 &0.14 & 5.2 & 0.05--0.09 \\
Ku-Band&13.7 & 3500&8.4 &14.0  &7.5 &0.2--0.3& 5.3 & 0.1 \\
KFPA &22 &8000 &6.7 & 17.3& 7.3&0.9--1.3 & 0.4  &0.1--0.2 \\
\hline
\end{tabular}
\caption{Table showing various parameters of the GBT receivers with corresponding survey limit for a future survey of the GC (see text for details). The details for receivers are given in \protect\url{https://science.nrao.edu/facilities/gbt/facilities/gbt/proposing/GBTpg.pdf}.
\label{tab:GCsurv}}
\end{table*}

\begin{figure*}
\centering
\includegraphics[scale=0.4]{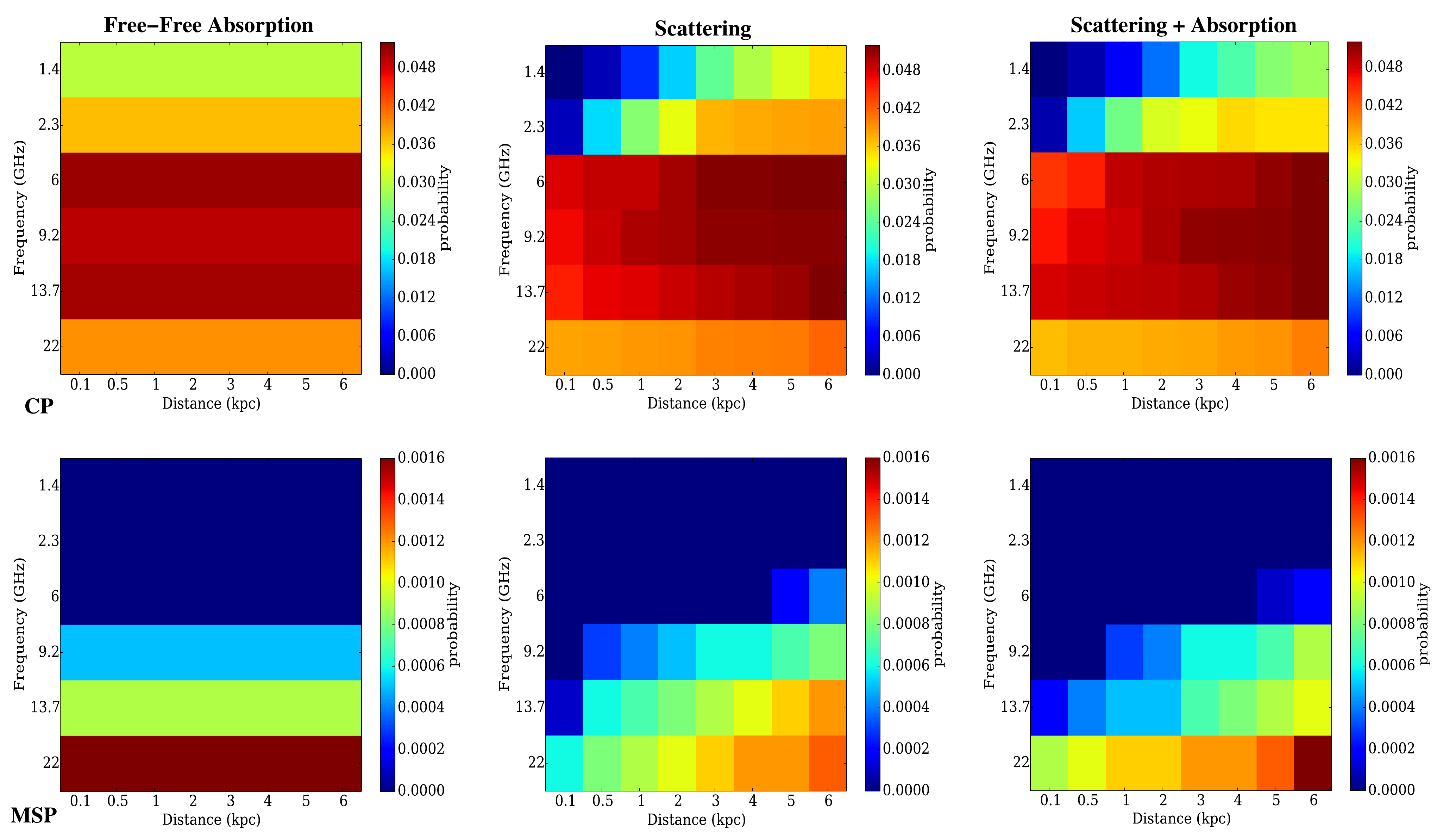}
%{\mbox{\psfig{file=B1133+16_simspec.eps,width=8cm}}}
\caption{Probability of finding a pulsar in the GC as function of frequency 
and distance of the scattering screen from the GC. The probabilities have been 
computed for future GBT surveys and assuming VEGAS as the backend. The banding seen in the free-free absorption case is due to the different bandwidths of receivers on the GBT.}
\label{fig:2dhist_v}
\end{figure*}

In 2012, a new backend was developed for the GBT. The VErsatile GBT 
Astronomical Spectrometer (VEGAS) is currently being used observations~\citep{Bu12}. The 
backend consists of 8 different spectrometer banks and has a maximum total 
instantaneous bandwidth of 1250~MHz for pulsar observations. VEGAS is 
expected to be the primary backend for pulsar astronomy and will replace 
the Green Bank Ultimate Pulsar Processing Instrument (GUPPI)~\citep{Ra08} in the process. Hence, in our analysis, we assume VEGAS to be the primary 
backend for future GBT pulsar surveys.
Under these assumptions, we computed probability of detection for two scenarios: (a) the backend would be able to accommodate the entire bandwidth of 
each receiver; (b) using VEGAS as the backend in which case the 
bandwidth is limited to 1250 MHz. The 2-D histograms for both the cases are shown in 
Fig.~\ref{fig:2dhist} and Fig.~\ref{fig:2dhist_v}.

\subsection{Results}
Tables~\ref{tab:survres} and~\ref{tab:GCsurv} clearly summarize our results 
from the analysis mentioned above. Table~\ref{tab:survres} shows the upper limits on the populations based on
previous and future surveys for various models. The results point out that based on the null results from previous surveys, we can obtain an upper limit on the CP and MSP population in the GC and the results do not reject an
existence of CP population in the GC. With the expected performance of
SKA-MID and ngVLA, we would be able to probe a sizable population of GC pulsars
which would give us much better constraints. The constraints on the pulsar population are less stringent as we include models for flux density mitigation as we would
detect a lesser fraction of the existing population due to the effects of the ISM. Table~\ref{tab:GCsurv} summarizes probabilities of finding one pulsar in a
potential GBT survey. Results show that CPs have a better prospect of being
detected than MSPs though the absolute probability is only as high as 0.07. Moreover, Table~\ref{tab:survres} suggests that current observations are less 
constraining on the MSP population than the CP population. The small number of predicted CPs would suggest that 
star formation is suppressed at the GC and that the existence of MSPs could be 
explained through capture of MSPs from globular cluster~\citep{Ho16}.

\section{Discussion}

Although the probability of detecting a single 
pulsar is greater than zero for higher frequencies, where scattering and absorption effects are 
negligible, the value itself is small.
This can be attributed to the distance of the GC where the flux densities of pulsars in the GC would be so small that even without assuming any attenuation 
of the flux density, we have been able to probe only a small fraction of the population. Irrespective of the dominance of sub populations in the GC (CP or MSP), the 
faintness of these sources due to the distance of the GC makes it difficult to 
detect them. This is clearly indicated by Fig.~\ref{fig:surv} where the survey 
sensitivity 
limit only encloses $0-2\%$ of the total simulated population of CPs and 0$\%$ of the total MSP population for the baseline model. This shows that we need deeper searches of the GC in the future even if the environment does not 
play a role in affecting pulsar flux densities. Our results allow for 
$\sim$ 445 CPs beaming towards us which is a less constraining compared to the results in~\cite{Ch14} by a factor of 2. We also obtain an upper limit of 5800 MSPs in the GC which is far less constraining compared to the CPs.~\cite{Ch14} take into account the magnetar population as a magnetar fraction in the GC and their results suggest previous surveys were not sensitive to existing pulsar population in the GC.~\cite{De14a} suggest 
that given the absence of hyperstrong scattering and lack of pulsar 
detections, there might be an intrinsic deficit of pulsars in the GC though 
our simulations suggest our radio surveys have not been sensitive enough to 
detect any pulsars in the GC. The detection of one magnetar hints at a 
preference to creation of magnetars in the GC. Future SKA and ngVLA surveys 
will be able to answer these questions.

Figs.~\ref{fig:2dhist} and~\ref{fig:2dhist_v} show the probability of 
detection for different frequencies and screen distances for MSPs and CPs. The figures show that free-free absorption
has negligible effect on the flux density mitigation beyond frequencies of 1~GHz due 
to negligible optical depths at higher frequencies. 
Hence, the probability of detection is solely dependent on the bandwidth of 
the telescope receivers. The banding structure evident in the figure is due to 
the fact that different GBT receivers have different bandwidths.
On the other hand, scattering plays an important role in reducing flux density from
pulsars. Scattering transitions from strong scattering to weak scattering 
regime as the distance of the screen from the GC increases. Hence, one would expect to have maximum yield from the GC survey when the screen is far enough from the GC and the survey is at a high frequency. These aforementioned effects 
help us in constraining the optimum frequency for future GC surveys. Note that 
the optimum frequency largely depends on the bandwidth of the survey if it is
backend limited. The 2-D histograms also suggest that the optimum frequency 
for future GBT surveys is as high as 9~GHz for CPs and 22~GHz for MSPs for 
strong and weak scattering cases if we assume the 
backends can cover the whole bandwidth of the receiver. On the other hand, if 
we consider VEGAS as the backend for future surveys, we obtain an optimum 
frequency of $\sim$9~GHz for CPs for both, the strong and the weak scattering 
case. For MSPs, the optimum frequency is 22~GHz for the weak and strong 
scattering case. Since we are interested in finding CPs and MSPs, based on these results, we propose that the optimal range of frequencies for future GBT surveys is 9--14~GHz. We 
also note that future surveys of the GC in the range of 1.4--6~GHz will not be able to detect MSPs, not because 
of absorption but the faintness of the sources. In any case, we have to go to higher frequencies ($>$~9~GHz) to detect any pulsars in the GC in single 
observational tracks.  

The results suggest that it would be more difficult to detect 
MSPs than CPs given the lower radio luminosities and the effect scattering has on their radio flux densities. We cannot favour any population at the moment because 
the analysis suggests that previous surveys have not been sensitive to any of 
the populations so far, even without factoring in the sources of flux density mitigation. Our conclusions differ from~\cite{Ma15}, which can be attributed to the fact that the population used
by~\citep{Ma15} is the actual pulsar population, which might have an inherent
selection bias in the pseudo-luminosity function of the source population as only the 
brightest pulsars have been detected by current radio telescopes. Hence, we sample only the tail of the underlying pseudo-luminosity distribution of pulsars, which can lead to different inferences about the source population. On the other hand, we have considered a
synthetic population of pulsars in the GC, assuming an underlying pseudo-luminosity
function, which properly accounts for this selection bias.  
 
Recent results suggest that scattering does not play an important role in the attenuation of flux densities towards the GC. This is an important result for 
future surveys of the GC. If the weak scattering scenario is true then 
Fig.~\ref{fig:surv} suggests that deeper searches of the GC without going to 
higher and higher frequencies would result in more detections of pulsars. 
Future telescopes like the SKA and next generation Very Large Array (ngVLA)~\citep{Hu15} will provide a great opportunity to search for radio pulsars in the GC. 
These surveys are expected to detect significant fraction of the pulsar 
population in the inner Galaxy. Future high frequency radio surveys with highly sensitive radio telescopes will help in resolving the pulsar problem in the GC.

\section{Conclusions}
In summary, from an analysis of the current observational constraints of the pulsar population in the GC, our main conclusions are as follows:
(i) the null results from previous surveys are not surprising, given that current surveys have only probed $\sim$~2$\%$ of the total CP population and 0$\%$ of the MSP population;
(ii) upper limits on the CP and MSP population for various models constrain the population of pulsars beaming towards us to be $<445$~CPs and $<5800$~MSPs;
(iii) a future GC survey with the GBT would have greater prospects of detecting CPs compared to MSPs. We find that the optimum frequency of a GBT survey would be 9--14~GHz;
(iv) a future surveys with SKA-MID and ngVLA would probe a sizable population of the pulsar population in the GC.

\section*{Acknowledgements}
We thank our anonymous referee for suggestions that vastly improved the manuscript. This research was partially supported by the National Science Foundation under Award No. OIA-1458952. Any opinions, findings and conclusions, or 
recommendations expressed in this material are those of the authors and do not 
necessarily reflect the views of the National Science Foundation.

\appendix

\section{Calculating reduction in flux due to scattering}

Here, we describe the method to calculate the reduction in flux due to
scattering.  Since pulsar surveys make use of harmonic summing to
increase the signal to noise of the detection in the Fourier domain,
for each of previous and future survey, we find the optimum number of
harmonics to be summed. For any survey, we follow the terminology
in~\cite{Co97} and define the ``pulsed fraction''
\begin{equation}
\eta_{p} = \sum_{0}^{N_{h}} \frac{R_{l}}{\sqrt{N_{h}}},
\label{eq:pf}
\end{equation} 
where $N_{h}$ is the number of harmonics to be summed and
\begin{equation}
R_{l} = \frac{S(l)}{S(0)}
\end{equation}
is the ratio of the amplitude of the $l^{th}$ harmonic and the
amplitude of the DC component in the Fourier domain.  For this
analysis, we assume a Gaussian pulse characterized by
\begin{equation}
f_{1}(t) = \frac{1}{\sqrt{2{\rm \pi}}\sigma}{\rm exp}\left[\frac{-t^{2}}{2\sigma^{2}} \right],
\end{equation}
where $\sigma$ is the standard deviation and in our
case, $t$ is time running over one pulse period, $P$.  For the
scattered case, we convolve the Gaussian with a one-sided exponential
function with a mean of $\tau_{s}$. This results in a modified pulse
profile described by
\begin{equation}
f_{2}(t) = \frac{\lambda}{2}~\exp{\left( \sigma^{2}\lambda - 
2t \right)}~{\rm erfc}\left( \sigma^{2}\lambda - t \right),
\end{equation}
where $\lambda=1/\tau_s$ and $\sigma$ is the
standard deviation of the Gaussian distribution and the complimentary
error function
\begin{equation}
  {\rm erfc}(x)=\frac{2}{\sqrt{\pi}} \int_x^{\infty} e^{-y^2} dy.
\end{equation}
The scattering broadening function in time domain is given by,
\begin{equation}
f_{sca}(t) = {\rm exp}\left(-\frac{t}{\tau_{s}}\right).
\end{equation}
In Fourier space, where the frequency of the
$l^{\rm th}$ harmonic $k=l/P$, the Gaussian pulse transforms to 
\begin{equation}
S_{\rm Gauss}(k) =\frac{1}{\sqrt{2\rm \pi}}{\rm exp}\left(-\frac{\sigma^{2}k^{2}}{2}\right)
\end{equation}
and the scatter broadening function transforms to
\begin{equation}
S_{\rm sca}(k) = \frac{1}{\left(k^{2}\tau_{s}^{2} + 1\right)}.
\end{equation}
The resulting Fourier components are then
\begin{equation}
S(k) = S_{\rm Gauss}(k) \cdot S_{\rm sca}(k).
\end{equation}
$f_{1}(t)$ and $f_{2}(t)$ reported here are already normalized to make sure that the area under the
pulse within one pulse period is the same for both functions. After
normalizing the pulse from both scenarios, we computed the Fourier
transform for the standard and scattered pulse.

\begin{figure*}
\centering
\includegraphics[scale=0.35]{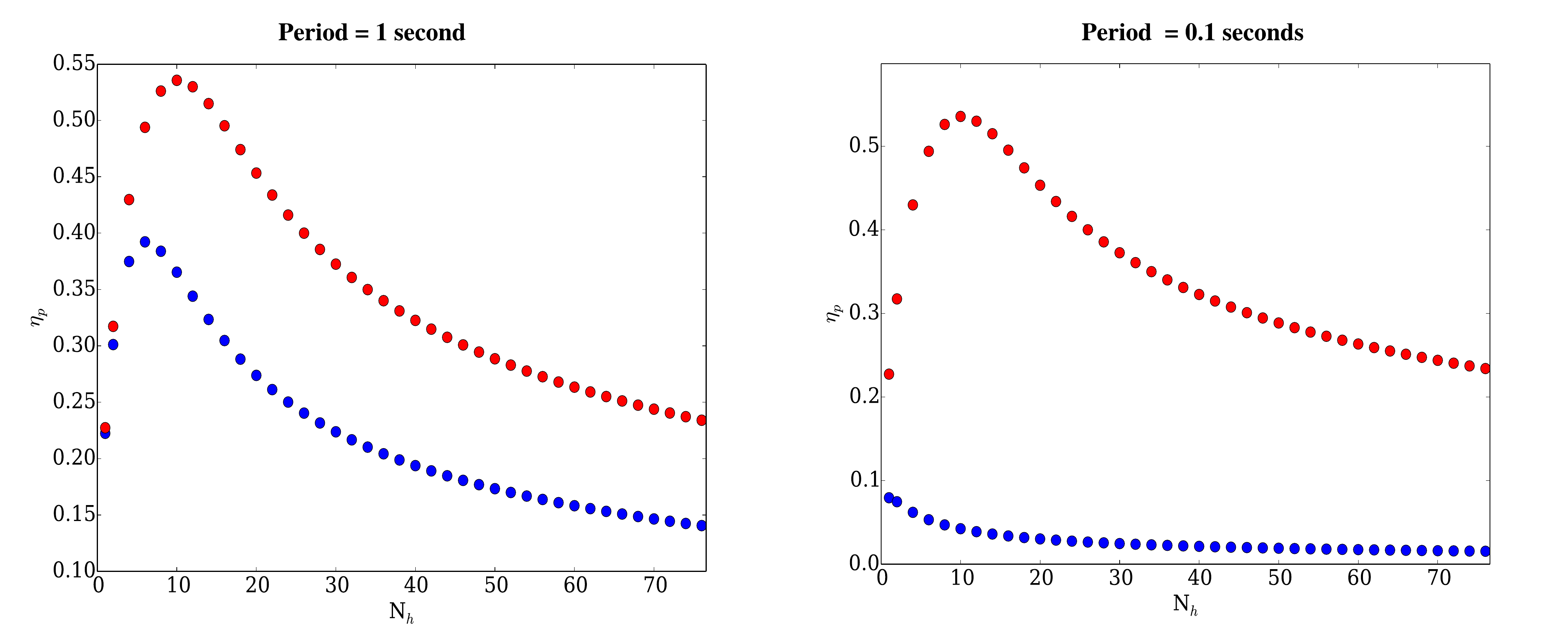}
\caption{Pulsed fraction versus number of harmonics summed for a Gaussian pulse (red dots) and a scattered pulse (blue dots). We assume a constant duty cycle of 5$\%$. We can see that smaller periods are severely affected by scattering. }
\label{fig:harsum}
\end{figure*}

Then, we obtained the optimal number of harmonics to be summed and
computed the pulsed fraction using Eq.~\ref{eq:pf}. The optimum number
of harmonics to be summed will be the value $N_{h}$ for which
Eq.~\ref{eq:pf} is maximized. Figure.~\ref{fig:harsum} shows one such
result for a strong scattering scenario for CPs for a fixed duty
cycle. In the case of strong scattering, the value of
$N_{h}$ is lower and the maximum value of the pulsed fraction is
significantly lower than the unscattered case. This means that the
sensitivity of the survey reduces by a factor of the ratio of the two
pulsed fractions,
\begin{equation}
\mathcal{S}(\nu) = \frac{\eta_{p,g}}{\eta_{p,sca}}.
\end{equation}

\bibliography{psrpop}{}

\bibliographystyle{mn2e}
\end{document}